\documentclass[aps,prl,preprint,groupedaddress]{revtex4-1}

\usepackage{amssymb}
\usepackage{graphics}

\begin{document}


\title{Signature of Cooper pairs in the non-superconducting phases of amorphous superconducting tantalum films}



\author{Yize Stephanie Li}

\affiliation{Department of Physics, University of Virginia, 
Charlottesville, VA 22904, USA}

\altaffiliation[Current Affiliation: ]{Department of Materials Science 
and Engineering, University of Wisconsin-Madison, 
Madison, WI 53706, USA}


\date{\today}

\begin{abstract}
We have studied the magnetic field or disorder induced insulating and 
metallic phases in amorphous Ta superconducting thin films. The evolution
of the nonlinear transport in the insulating phase exhibits a 
non-monotonic behavior as the magnetic field is increased. We suggest that 
this observation could be an evidence of the presence of localized Cooper 
pairs in the insulating phase. As the metallic phase 
intervenes the superconducting and insulating states in Ta films, this 
result further reveals that Cooper pairs also exist in the metallic ground state.
\end{abstract}


\maketitle

Superconductivity in homogeneously disordered two dimensional (2D) system 
is of particular interest because 2D is the lower critical dimension for 
both superconductivity and localization. Conventional theory predicts that 
in 2D the disorder- or magnetic field (B)- induced suppression of 
superconductivity leads to a direct superconductor-insulator transition 
(SIT) \cite{Fisher1, Fisher2, Finkelshtein, Larkin, Dubi} in the zero 
temperature ($T\rightarrow0$) limit. The ``dirty boson'' model 
 \cite{Fisher1, Fisher2}, which describes the superconducting 
phase as a condensate of 
Cooper pairs with localized vortices and the insulating phase as a 
condensate of vortices with localized Cooper pairs, assumes the presence 
of Cooper pairs on both sides of the SIT. The existence of localized Cooper 
pairs in the insulating phase of granular films is undoubted, however, 
whether Cooper pairs are present in the insulating phase of amorphous 
films is a more complicated issue \cite{Goldman}. Experimental evidence on 
the crossover from Bose insulator with nonzero pairing to Fermi electronic 
insulator without pairing in amorphous InO films \cite{Paalanen}, and the emergence 
of a magnetoresistance (MR) peak in several 2D amorphous materials 
\cite{Paalanen, Hebard, sGantmakher2, Sambandamurthy1, 
Steiner, Okuma1, Okuma2, Parendo}, such as InO \cite{Paalanen, 
Hebard, sGantmakher2, Sambandamurthy1, Steiner}, 
MoSi \cite{Okuma1, Okuma2}, and Bi \cite{Parendo}, and the observation of 
activated resistances and magnetoresistance oscillations dictated by the 
superconducting flux quantum in patterned amorphous Bi films \cite{Stewart}, 
suggest that Cooper pairs might be present in the insulating phase of homogeneously 
disordered superconducting films. More recently, the direct evidence for the 
existence of preformed Cooper pairs in non-superconducting states of amorphous 
InO \cite{sSacepe2}, TiN \cite{sSacepe1}, and NbN \cite{sChand} films is 
revealed by scanning tunneling spectroscopy. However, question concerning 
whether the persistence of Cooper pairs in insulating phase is a generic or 
a material-based property of amorphous superconducting films remains.  

While the nature of the insulating state is still an issue to be solved, 
the subject has attracted more attention because of the observation of the 
metallic phase in amorphous MoGe \cite{EYKB, Mason} and Ta 
\cite{Yoon2, Yoon1, Yoon3, Yize1} thin films under 
weak magnetic fields. The unexpected metallic behavior, which intervenes 
the B- or disorder- driven SIT, is characterized by a drop in resistance 
($\rho$) followed by a saturation to a finite value as $T\rightarrow0$.
This low field- or disorder- driven metallic phase is significantly 
different from the high field ``quantum metal'' observed in
insulating Be films \cite{Butko} or superconducting InO films 
\cite{sGantmakher2} and TiN films \cite{Baturina}, which is induced by 
high B fields on films that already exhibit insulating phase. 
Several theoretical models have been proposed to account for the emergence 
of the metallic ground state \cite{DP, WP, Galitski, SAK, SOK}, 
including the quantum phase glass model \cite{DP, WP}, the quantum vortex 
picture \cite{Galitski}, and the percolation paradigm \cite{SAK, SOK}. However, a consensus 
on the mechanism for the metallic behavior hasn't been reached yet. 

In this work, we study the evolution of nonlinear 
current-voltage (I-V) characteristics in the insulating phase of amorphous 
superconducting Ta films with the increase of B field. The non-monotonic B 
field dependence of the dV/dI peak, which has been observed in all the films 
we have studied, suggests the existence of the localized Cooper pairs in 
both B- and disorder- driven insulating phases. Our study also implies that 
the nonlinear transport characteristics might be a consistent and sensitive 
probe to detect localized Cooper pairs. 

Our Ta thin films are dc sputter deposited on Si substrate and are
patterned into a 1mm wide and 5mm long Hall bar for the standard four point
measurement using a shadow mask. The thickness of the films is between 
6 nm and 2 nm. As reported before \cite{Yoon2}, the x-ray
diffraction pattern of such films does not show any sign of local atomic correlation,
so they are structurally amorphous and homogeneously disordered, as expected
from the excellent wetting property of Ta on almost all substrates. The temperature
dependence of resistivity measurement indicates that the superconducting transition 
temperature decreases continuously toward 0 K with increasing disorder and the 
transition exhibits no reentrant behavior, which further confirms the 
amorphous nature of these Ta films \cite{Goldman}. The magnetoresistance was measured 
by a lock-in with 1 nA ac current. The dV/dI trace was measured by a homemade ac+dc 
currents summing circuit which used a lock-in to modulate the dc bias current with 
a small ac amplitude at low frequency. 

\begin{table}
\caption{\label{table1}List of sample parameters: nominal film thickness 
t, normal state resistivity $\rho_{n}$ at 4.2 K, and the observed 
phases at low temperature (S for the superconducting phase, M for 
the metallic phase, and I for the insulating phase). For samples 
exhibiting the superconducting phase, we list mean field $T_{c}$ at B = 0, 
the critical magnetic field $H_{c}$ as defined by the field at which the 
low temperature (60 mK for Ta 1 - Ta 3, 50 mK for Ta 4) resistance 
reaches 90\% of the high field saturation value, and the superconducting 
coherence length calculated from $\xi = \sqrt{\Phi_{0} / 2\pi B_{c}}$ , 
where $\Phi_{0}$ is the flux quantum.} 
\begin{ruledtabular} 
\begin{tabular}{cclcrccc}
Films & Batch & t(nm) & $\rho_{n}$(k$\Omega / \square$) & phase & 
$T_{c}$(K) & $H_{c}$(T) & $\xi$(nm) \\
Ta 1  & 1 & 5.6  & 1.42 & S,M,I & 0.65 & 0.82 & 20 \\
Ta 2  & 1 & 4.6  & 1.85 & S,M,I & 0.54 & 0.68 & 22 \\
Ta 3  & 1 & 5.1  & 2.16 & S,M,I & 0.38 & 0.58 & 24 \\
Ta 4  & 2 & 4.1  & 2.28 & S,M,I & 0.26 & 0.33 & 32 \\
Ta 5  & 3 & 2.8  & 4.62 &   M,I &      &      &    \\
Ta 6  & 4 & 2.5  & 6.24 &     I &      &      &    \\
Ta 7  & 4 & 2.5  & 8.00 &     I &      &      &    \\
Ta 8  & 3 & 2.36 & 8.78 &     I &      &      &    \\
\end{tabular}
\end{ruledtabular}
\end{table}

Recent studies on Ta films have shown that each phase displays remarkably 
different nonlinear I-V characteristics \cite{Yoon2}, offering an alternative 
criterion to identify phases which is fully consistent with that based on the 
T dependence of $\rho$. The superconducting phase is unique in exhibiting 
hysteresis in the I-V curve, which has been demonstrated to arise from a 
nonthermal origin \cite{Yoon1}. As the system is driven into the metallic 
phase, the hysteresis evolves into the point of the largest slope in the 
continuous and reversible I-V implying that the nonlinear transport in 
the metallic phase ($d^2V/dI^2 > 0$) is also intrinsic and uncorrelated with 
electron heating effect \cite{Yoon1}. The I-V characteristics and accompanying 
long electronic relaxation time in the superconducting and metallic phases can 
be well explained by the vortex dynamics picture \cite{Yize1}. The insulating 
phase is characterized by a peak structure in the dV/dI vs. I trace ($d^2V/dI^2 < 0$),
and has been employed as a phenomenological symbol to identify the phase \cite{Yize1}.

Qualitatively similar dV/dI peaks have been observed in the insulating phases of 
TiN \cite{sVinokur}, InO \cite{sSambandamurthy2}, and MoGe \cite{Yazdani}.
However, the origin of the insulating nonlinear transport might be different 
for various systems. The nonlinear I-V in strong insulators, especially 
the giant jumps of current at finite voltages in InO and TiN \cite{Ovadia, 
sVinokur}, is believed to be a consequence of overheating of electrons 
\cite{Altshuler}. We note that the amorphous Ta films we studied are 
significantly different from those systems. Firstly, the I-V characteristic 
in the insulating phase of Ta is continuous showing no sign of a bistability. 
Secondly, the validity of the proposed electrons overheating model \cite{Altshuler} 
requires a steep temperature dependence of the resistance, which is the case 
for InO and TiN but not the case for Ta \cite{Yize1}. Furthermore, the 
non-monotonic B field dependence of the dV/dI peak structure in the insulating
phase of Ta, especially the fact that the most pronounced non-monotonic 
feature appears in the B-induced insulating phase of low disordered 
films which have lower resistance and thus less likely to be overheated 
compared with the highly disordered films, as presented in this paper, 
indicates that the electron heating effect, if plays a role, could not be the only 
source for the nonlinear transport in the insulating phase of Ta films.     

\begin{figure}
\includegraphics{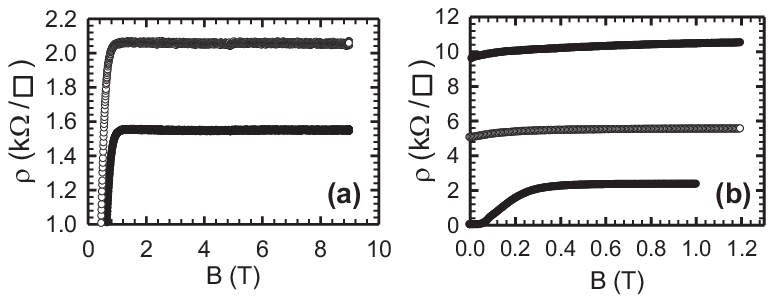}
\caption{\label{fig1}(a) Magnetoresistance for Ta 1 (bottom trace) and
Ta 2 (top trace) at T = 60 mK. (b) Magnetoresistance for (from bottom
to top) Ta 4, Ta 5, and Ta 7 at T = 50 mK.}
\end{figure}

Conventionally, the B-induced insulating phase was studied by MR 
measurements \cite{Paalanen, Hebard, sGantmakher2, 
Sambandamurthy1, Steiner, Okuma1, Okuma2, Parendo}.  The negative MR 
is often interpreted as a consequence of a conduction enhancement due to 
delocalization and/or breaking of the localized Cooper pairs which are 
believed to be present in the insulating phase \cite{Fisher1, Fisher2}. The positive MR is usually 
attributed to the response of unpaired electrons \cite{Okuma1, Okuma2, Parendo, sMatveev, sMertes}. Figure 1 
shows the MR measured on five Ta films with different degrees of disorder. 
Parameters of the films are summarized in Table I. The resistivity of each 
film increases monotonically with the magnetic field and eventually 
saturates, implying that the localized Cooper pairs in the insulating phase of 
Ta films, if present, are not detectable in the linear transport regime. 

Figure 2(a) illustrates the evolution of differential I-V 
in the B-induced metallic and insulating phases in superconducting 
film Ta 1. For B $\leq$ 0.8 T (B $\geq$ 0.9 T), the dV/dI is a 
monotonically increasing (decreasing) function of the bias current,
which characterizes the metallic (insulating) phase.  
At B = 0.85 T (thick solid line), the sign of $d^{2}V/dI^{2}$ is 
positive (negative) at high (low) bias currents as in the metallic 
(insulating) phase at lower (higher) fields. The non-monotonic dV/dI 
with respect to bias current was interpreted as an evidence that the 
insulating state near $B_{c}$ consists of metallic domains connected 
by point contacts \cite{Yoon3}. Fig. 2(b) shows the 
MR measured in the low current limit (1 nA) at three 
different temperatures within the low T regime. The presence of a 
crossing point at $B_{c}$ = 0.83 T, which is defined as the critical field 
for metal-insulator transition, confirms that the transport at low currents 
at B = 0.85 T is insulating. 

\begin{figure}
\includegraphics{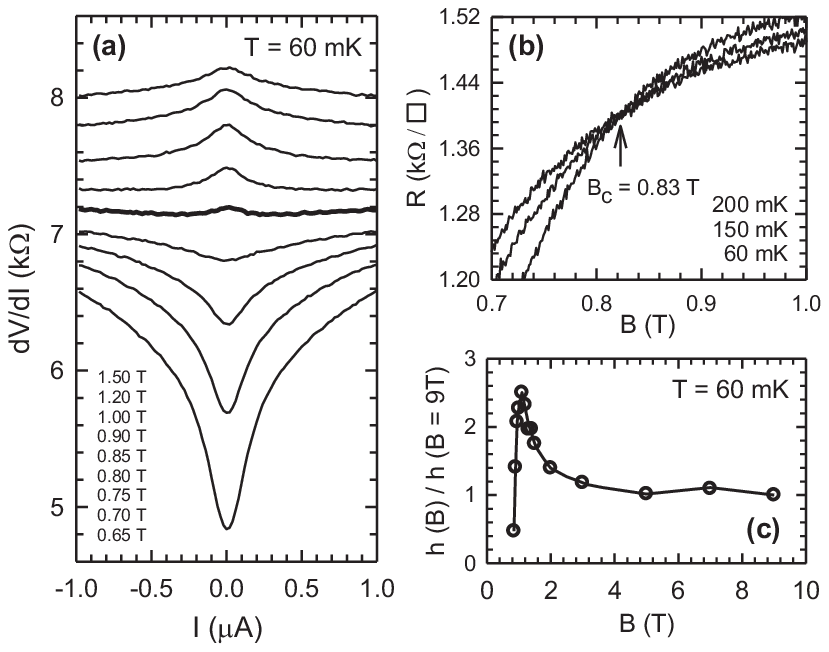}
\caption{\label{fig2}(a) Evolution of dV/dI vs. I with increasing B for 
Ta 1 at T = 60 mK. Each trace is vertically shifted successively for clarity. 
(b) Magnetoresistance for Ta 1 at the indicated temperatures (from top to bottom 
below the crossing point). The crossing point at 0.83 T marked by an arrow defines 
the critical field $B_{c}$ for metal-insulator transition. (c) Normalized height 
of the dV/dI peak structure h(B)/h(B=9T) vs. B for Ta 1 at T = 60 mK. Symbols are 
experimental data and solid lines are to guide the eye. }
\end{figure}

In this work, we focus on the evolution of dV/dI peak structure in the insulating 
phase with the increase of magnetic field. The height of the dV/dI peak in 
Fig. 2(a) initially increases with increasing B, reaching a maximum at B = 1.1 T, 
and then decreases with further increase of B. We use the normalized peak height 
h(B)/h(B=9T) as a quantitative measure of the non-monotonic feature we observe and
plot it as a funtion of the magnetic field in Fig. 2(c). h(B)/h(B=9T) increases 
with B field until reaching a prominent maximum and then decreases, showing a 
general trend of saturation. In the context of ``dirty boson'' model, the dV/dI peak 
is attributed to the current-induced delocalization of the localized Cooper 
pairs \cite{Yoon2}. We thus expect that the dV/dI peak height initially grows 
as the system is driven into the insulating phase where Cooper pairs are localized. 
Above a certain magnetic field ($\sim$ 1.1 T for Ta 1), however, the population of the 
localized Cooper pairs would decrease with increasing B because of the B-induced 
pair breaking mechanism. As the population of the localized Cooper pairs decreases, 
the effect of their current-induced delocalization would consequently decrease,
resulting in a reduced dV/dI peak height. We suggest that this non-monotonic feature is a result of 
superconducting correlations. Although the electron heating effect and/or Coulomb interaction between 
normal electrons might also lead to dV/dI peak structures, neither of them 
could be responsible for the observed non-monotonic evolution of the peak structure 
as the magnetic field is increased. This result is thus interpreted as a signature of the 
localized Cooper pairs in the B-induced insulating phase of superconducting Ta films. 

\begin{figure}
\includegraphics{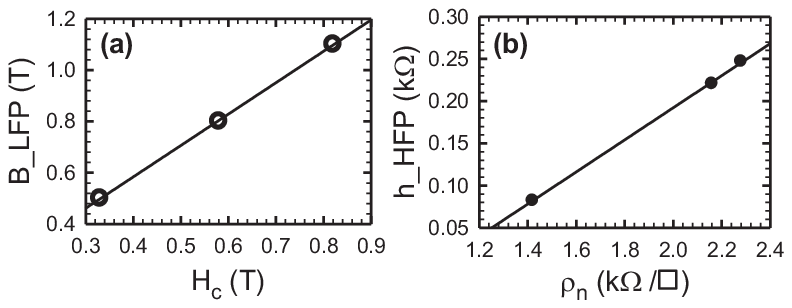}
\caption{\label{fig3}(a) The magnetic field where the low field maximum occurs in the
h(B)/h(B=9T) vs. B trace as a function of the critical magnetic field $H_{c}$ for
3 low disordered films, Ta 1, Ta 3, and Ta 4. (b) The height of the dV/dI peak where
the high field maximum takes place as a function of the normal state resistivity $\rho_{n}$ 
for Ta 1, Ta 3, and Ta 4. Symbols are experimental data and solid lines are linear fits. }
\end{figure}

Such non-monotonic feature in the dV/dI peak height vs. B plot is 
evident for all low disordered Ta films we have studied. Figure 3(a)
shows the B field where the maximum occurs as a function of the critical magnetic
field $H_{c}$ for Ta 1, Ta 3, and Ta 4. The data falls into a line in an almost 
perfect fashion. As $H_{c}$ is directly related to the superconducting coherence 
length $\xi$, this result further suggests the intrinsic link between the non-monotonic
feature and superconducting correlations. In addition to the prominent maximum
in the dV/dI peak height vs. B plot at $\sim$ 1 T or below, a shallow peak
at $\sim$ 7 T, as shown in Fig. 2(c) is also observed in the B-induced insulating 
phase of all superconducting films. The origin for this high field feature,
which is most likely not due to superconductivity related mechanism \cite{Klich, Dalidovich}, 
is not clear at present and might require systematical studies at higher magnetic fields.
To distinguish the low field non-monotonic feature from the high field maximum,
we name the former as low field peak (LFP) and the latter as high field peak (HFP).
The height of the HFP is found to show a linear dependence on the normal state
resistivity, as shown in Fig. 3(b).

In addition to low disordered samples which exhibit superconducting behaviors 
at low T and low B, we also studied the evolution of dV/dI as a function of B field 
for highly disordered films that are insulating at low T at B = 0. Figure 4 shows 
dV/dI spectra and normalized peak height vs. B for two insulating samples. A low
field maximum of the dV/dI peak height, which is less pronounced than that in 
Fig. 2, emerges at $\sim$ 1 T as shown in Fig. 4(a) and 4(b) for Ta 7. Figure 4(c) 
shows the dV/dI spectra of Ta 8, a sample with a higher degree of disorder than Ta 7. 
As indicated in Fig. 4(d), a fairly weak maximum appears at $\sim$ 1 T in the dV/dI 
peak height vs. B plot. The weakening of the B-induced Cooper pairs breaking effect 
with the increase of disorder, could be a consequence of a reduced population of 
Cooper pairs as the degree of disorder is increased. Alternatively, it might be 
caused by smearing of the B-induced Cooper pairs breaking effect by disorder \cite{Dalidovich}, 
or by electron heating which, if contributes to the dV/dI peak structures, would 
play an increasingly more important role with the increase of disorder.   

\begin{figure}
\includegraphics{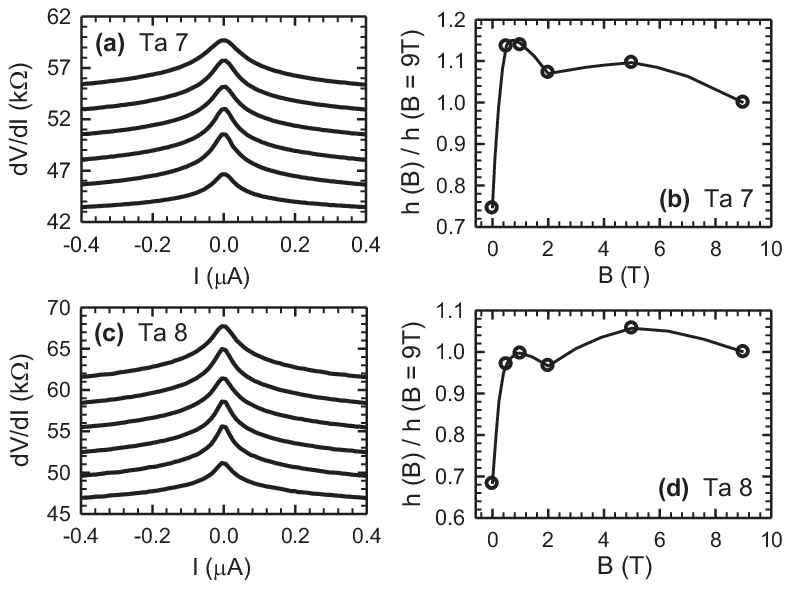}
\caption{\label{fig4}(a) Evolution of dV/dI vs. I for Ta 7 at T = 60 mK
at B = 0 T, 0.5 T, 1 T, 2 T, 5 T, and 9 T. Each trace is vertically
shifted successively. (b) Normalized height h(B)/h(B=9T) vs. B for Ta 7
at T = 60 mK. Solid lines are to guide the eye. (c) Evolution of dV/dI 
traces (vertically shifted) for Ta 8 at the same temperature and magnetic
fields as in panel (a). Ta 8 is more disordered than Ta 7 by the measure 
of $\rho_{n}$. (d) Normalized height as a function of B field for Ta 8 
at T = 60 mK. Solid lines are to guide the eye.}
\end{figure}

A high field maximum is also observed in these highly disordered insulating films, 
as shown in Fig. 4(b) and 4(d), which probably has the same origin as the HFP
for low disordered samples. The heigths of the dV/dI peaks at 1 T 
(open symbols) and 5 T (filled symbols), and their ratio are plotted as a 
function of $\rho_{n}$ in Fig. 5, for 3 highly disordered films, Ta 6 - Ta 8. The peak 
heights grow with the increase of disorder, and a linear dependence on $\rho_{n}$
is observed for the peak at B = 1 T. The ratio of the peak heights at 1 T and 5 T 
decreases from above 1 to below 1, as $\rho_{n}$ is increased, indicating that the 
LFP is more sensitive to the disorder.  
   
\begin{figure}
\includegraphics{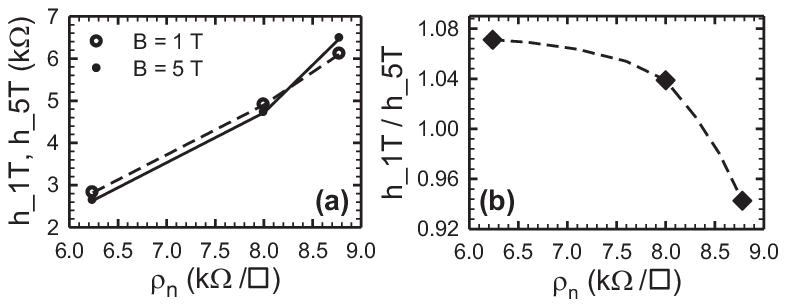}
\caption{\label{fig5}(a) dV/dI peak height at B = 1 T (open symbols) and 5 T (filled
symbols) vs. normal state resistivity $\rho_{n}$ for 3 highly disordered films, Ta 6 - Ta 8.
(b) Ratio of the dV/dI peak heights at B = 1 T and 5 T as a function of $\rho_{n}$ for
Ta 6 - Ta 8. Solid and dashed lines are to guide the eye.}
\end{figure}

The evolution of dV/dI peak structure with the increase of B field for films 
with various degrees of disorder, as shown in Fig. 2 - Fig. 5, suggests that Cooper 
pairs exist in both B- and disorder- induced insulating phases of Ta 
films and can be detected by nonlinear electronic transport. Although this result
could not exclude the possible contributions of unpaired normal electrons and/or 
electron heating to the dV/dI peak in the insulating phase, neither of them could
produce the observed low-field non-monotonic feature, which would not be possible
without the presence of Cooper pairs. In a recent paper \cite{Yize1}, 
we mapped the phase diagram of Ta thin films in B-T-disorder 
space, which indicates that the superconducting phase is completely 
surrounded by the metallic phase prohibiting a direct SIT at any disorder. 
Thus our result further reveals the presence of Cooper pairs in the metallic phase 
which intervenes the superconducting and insulating regimes, as proposed 
by Phillips {\em et al} \cite{DP, WP}.  

Although a lot of theoretical efforts have been devoted to understand 
the transport properties of disordered 2D superconductors 
in the linear current response regime, not much theoretical work has been carried 
out to study the nonlinear I-V. One of the major reasons might be that very 
limited experimental results in the nonlinear current response regime have 
been reported. The work presented here would stimulate the development of 
more comprehensive theoretical models to account for the transport behaviors 
in the nonlinear I-V regime of disordered superconducting films. 

To summarize, we have reported that in the insulating transport regime of 
homogeneously disordered Ta films, the height of the dV/dI peak structure 
experiences a non-monotonic change with the increase of magnetic field. 
This observation suggests that Cooper pairs persist into the insulating 
phase of the films, and further implies the presence of Cooper pairs in 
the metallic phase which intervenes both the B-induced and the disorder -induced 
superconductor-insulator transition. Our work indicates that, compared 
with the traditional magnetoresistance measurement, nonlinear transport 
characteristics might be a more sensitive probe of localized Cooper pairs.

The author thanks Jongsoo Yoon, Bascom Deaver, and Robert Weikle for providing 
experimental resources and discussions, and thank Stuart Wolf, Israel Klich,
and Denis Dalidovich for discussions.

\bibliography{SSTSep2014}

\begin{thebibliography}{40}%
\makeatletter
\providecommand \@ifxundefined [1]{%
 \@ifx{#1\undefined}
}%
\providecommand \@ifnum [1]{%
 \ifnum #1\expandafter \@firstoftwo
 \else \expandafter \@secondoftwo
 \fi
}%
\providecommand \@ifx [1]{%
 \ifx #1\expandafter \@firstoftwo
 \else \expandafter \@secondoftwo
 \fi
}%
\providecommand \natexlab [1]{#1}%
\providecommand \enquote  [1]{``#1''}%
\providecommand \bibnamefont  [1]{#1}%
\providecommand \bibfnamefont [1]{#1}%
\providecommand \citenamefont [1]{#1}%
\providecommand \href@noop [0]{\@secondoftwo}%
\providecommand \href [0]{\begingroup \@sanitize@url \@href}%
\providecommand \@href[1]{\@@startlink{#1}\@@href}%
\providecommand \@@href[1]{\endgroup#1\@@endlink}%
\providecommand \@sanitize@url [0]{\catcode `\\12\catcode `\$12\catcode
  `\&12\catcode `\#12\catcode `\^12\catcode `\_12\catcode `\%12\relax}%
\providecommand \@@startlink[1]{}%
\providecommand \@@endlink[0]{}%
\providecommand \url  [0]{\begingroup\@sanitize@url \@url }%
\providecommand \@url [1]{\endgroup\@href {#1}{\urlprefix }}%
\providecommand \urlprefix  [0]{URL }%
\providecommand \Eprint [0]{\href }%
\providecommand \doibase [0]{http://dx.doi.org/}%
\providecommand \selectlanguage [0]{\@gobble}%
\providecommand \bibinfo  [0]{\@secondoftwo}%
\providecommand \bibfield  [0]{\@secondoftwo}%
\providecommand \translation [1]{[#1]}%
\providecommand \BibitemOpen [0]{}%
\providecommand \bibitemStop [0]{}%
\providecommand \bibitemNoStop [0]{.\EOS\space}%
\providecommand \EOS [0]{\spacefactor3000\relax}%
\providecommand \BibitemShut  [1]{\csname bibitem#1\endcsname}%
\let\auto@bib@innerbib\@empty
\bibitem [{\citenamefont {Fisher}(1990)}]{Fisher1}%
  \BibitemOpen
  \bibfield  {author} {\bibinfo {author} {\bibfnamefont {M.~P.~A.}\
  \bibnamefont {Fisher}},\ }\href@noop {} {\bibfield  {journal} {\bibinfo
  {journal} {Phys. Rev. Lett.}\ }\textbf {\bibinfo {volume} {65}},\ \bibinfo
  {pages} {923} (\bibinfo {year} {1990})}\BibitemShut {NoStop}%
\bibitem [{\citenamefont {Fisher}\ \emph {et~al.}(1990)\citenamefont {Fisher},
  \citenamefont {Grinstein},\ and\ \citenamefont {Girvin}}]{Fisher2}%
  \BibitemOpen
  \bibfield  {author} {\bibinfo {author} {\bibfnamefont {M.~P.~A.}\
  \bibnamefont {Fisher}}, \bibinfo {author} {\bibfnamefont {G.}~\bibnamefont
  {Grinstein}}, \ and\ \bibinfo {author} {\bibfnamefont {S.~M.}\ \bibnamefont
  {Girvin}},\ }\href@noop {} {\bibfield  {journal} {\bibinfo  {journal} {Phys.
  Rev. Lett.}\ }\textbf {\bibinfo {volume} {64}},\ \bibinfo {pages} {587}
  (\bibinfo {year} {1990})}\BibitemShut {NoStop}%
\bibitem [{\citenamefont {Finkelshtein}(1987)}]{Finkelshtein}%
  \BibitemOpen
  \bibfield  {author} {\bibinfo {author} {\bibfnamefont {A.}~\bibnamefont
  {Finkelshtein}},\ }\href@noop {} {\bibfield  {journal} {\bibinfo  {journal}
  {JETP Lett.}\ }\textbf {\bibinfo {volume} {45}},\ \bibinfo {pages} {46}
  (\bibinfo {year} {1987})}\BibitemShut {NoStop}%
\bibitem [{\citenamefont {Larkin}(1999)}]{Larkin}%
  \BibitemOpen
  \bibfield  {author} {\bibinfo {author} {\bibfnamefont {A.}~\bibnamefont
  {Larkin}},\ }\href@noop {} {\bibfield  {journal} {\bibinfo  {journal} {Ann.
  Phys.}\ }\textbf {\bibinfo {volume} {8}},\ \bibinfo {pages} {785} (\bibinfo
  {year} {1999})}\BibitemShut {NoStop}%
\bibitem [{\citenamefont {Dubi}\ \emph {et~al.}(2007)\citenamefont {Dubi},
  \citenamefont {Meir},\ and\ \citenamefont {Avishai}}]{Dubi}%
  \BibitemOpen
  \bibfield  {author} {\bibinfo {author} {\bibfnamefont {Y.}~\bibnamefont
  {Dubi}}, \bibinfo {author} {\bibfnamefont {Y.}~\bibnamefont {Meir}}, \ and\
  \bibinfo {author} {\bibfnamefont {Y.}~\bibnamefont {Avishai}},\ }\href@noop
  {} {\bibfield  {journal} {\bibinfo  {journal} {Nature}\ }\textbf {\bibinfo
  {volume} {449}},\ \bibinfo {pages} {876} (\bibinfo {year}
  {2007})}\BibitemShut {NoStop}%
\bibitem [{\citenamefont {Goldman}\ and\ \citenamefont
  {Markovic}(1998)}]{Goldman}%
  \BibitemOpen
  \bibfield  {author} {\bibinfo {author} {\bibfnamefont {A.~M.}\ \bibnamefont
  {Goldman}}\ and\ \bibinfo {author} {\bibfnamefont {N.}~\bibnamefont
  {Markovic}},\ }\href@noop {} {\bibfield  {journal} {\bibinfo  {journal}
  {Phys. Today}\ }\textbf {\bibinfo {volume} {51(11)}},\ \bibinfo {pages} {39}
  (\bibinfo {year} {1998})}\BibitemShut {NoStop}%
\bibitem [{\citenamefont {Paalanen}\ \emph {et~al.}(1992)\citenamefont
  {Paalanen}, \citenamefont {Hebard},\ and\ \citenamefont {Ruel}}]{Paalanen}%
  \BibitemOpen
  \bibfield  {author} {\bibinfo {author} {\bibfnamefont {M.~A.}\ \bibnamefont
  {Paalanen}}, \bibinfo {author} {\bibfnamefont {A.~F.}\ \bibnamefont
  {Hebard}}, \ and\ \bibinfo {author} {\bibfnamefont {R.~R.}\ \bibnamefont
  {Ruel}},\ }\href@noop {} {\bibfield  {journal} {\bibinfo  {journal} {Phys.
  Rev. Lett.}\ }\textbf {\bibinfo {volume} {69}},\ \bibinfo {pages} {1604}
  (\bibinfo {year} {1992})}\BibitemShut {NoStop}%
\bibitem [{\citenamefont {Hebard}\ and\ \citenamefont
  {Paalanen}(1990)}]{Hebard}%
  \BibitemOpen
  \bibfield  {author} {\bibinfo {author} {\bibfnamefont {A.~F.}\ \bibnamefont
  {Hebard}}\ and\ \bibinfo {author} {\bibfnamefont {M.~A.}\ \bibnamefont
  {Paalanen}},\ }\href@noop {} {\bibfield  {journal} {\bibinfo  {journal}
  {Phys. Rev. Lett.}\ }\textbf {\bibinfo {volume} {65}},\ \bibinfo {pages}
  {927} (\bibinfo {year} {1990})}\BibitemShut {NoStop}%
\bibitem [{\citenamefont {Gantmakher}\ \emph {et~al.}(2000)\citenamefont
  {Gantmakher} \emph {et~al.}}]{sGantmakher2}%
  \BibitemOpen
  \bibfield  {author} {\bibinfo {author} {\bibfnamefont {V.~F.}\ \bibnamefont
  {Gantmakher}} \emph {et~al.},\ }\href@noop {} {\bibfield  {journal} {\bibinfo
   {journal} {JETP Lett.}\ }\textbf {\bibinfo {volume} {71}},\ \bibinfo {pages}
  {473} (\bibinfo {year} {2000})}\BibitemShut {NoStop}%
\bibitem [{\citenamefont {Sambandamurthy}\ \emph {et~al.}(2004)\citenamefont
  {Sambandamurthy}, \citenamefont {Engel}, \citenamefont {Johansson},\ and\
  \citenamefont {Shahar}}]{Sambandamurthy1}%
  \BibitemOpen
  \bibfield  {author} {\bibinfo {author} {\bibfnamefont {G.}~\bibnamefont
  {Sambandamurthy}}, \bibinfo {author} {\bibfnamefont {L.~W.}\ \bibnamefont
  {Engel}}, \bibinfo {author} {\bibfnamefont {A.}~\bibnamefont {Johansson}}, \
  and\ \bibinfo {author} {\bibfnamefont {D.}~\bibnamefont {Shahar}},\
  }\href@noop {} {\bibfield  {journal} {\bibinfo  {journal} {Phys. Rev. Lett.}\
  }\textbf {\bibinfo {volume} {92}},\ \bibinfo {pages} {107005} (\bibinfo
  {year} {2004})}\BibitemShut {NoStop}%
\bibitem [{\citenamefont {Steiner}\ \emph {et~al.}(2005)\citenamefont
  {Steiner}, \citenamefont {Boebinger},\ and\ \citenamefont
  {Kapitulnik}}]{Steiner}%
  \BibitemOpen
  \bibfield  {author} {\bibinfo {author} {\bibfnamefont {M.~A.}\ \bibnamefont
  {Steiner}}, \bibinfo {author} {\bibfnamefont {G.}~\bibnamefont {Boebinger}},
  \ and\ \bibinfo {author} {\bibfnamefont {A.}~\bibnamefont {Kapitulnik}},\
  }\href@noop {} {\bibfield  {journal} {\bibinfo  {journal} {Phys. Rev. Lett.}\
  }\textbf {\bibinfo {volume} {94}},\ \bibinfo {pages} {107008} (\bibinfo
  {year} {2005})}\BibitemShut {NoStop}%
\bibitem [{\citenamefont {Okuma}\ \emph {et~al.}(2001)\citenamefont {Okuma},
  \citenamefont {Shinozaki},\ and\ \citenamefont {Morita}}]{Okuma1}%
  \BibitemOpen
  \bibfield  {author} {\bibinfo {author} {\bibfnamefont {S.}~\bibnamefont
  {Okuma}}, \bibinfo {author} {\bibfnamefont {S.}~\bibnamefont {Shinozaki}}, \
  and\ \bibinfo {author} {\bibfnamefont {M.}~\bibnamefont {Morita}},\
  }\href@noop {} {\bibfield  {journal} {\bibinfo  {journal} {Phys. Rev. B}\
  }\textbf {\bibinfo {volume} {63}},\ \bibinfo {pages} {054523} (\bibinfo
  {year} {2001})}\BibitemShut {NoStop}%
\bibitem [{\citenamefont {Okuma}\ \emph {et~al.}(1998)\citenamefont {Okuma},
  \citenamefont {Terashima},\ and\ \citenamefont {Kukubo}}]{Okuma2}%
  \BibitemOpen
  \bibfield  {author} {\bibinfo {author} {\bibfnamefont {S.}~\bibnamefont
  {Okuma}}, \bibinfo {author} {\bibfnamefont {T.}~\bibnamefont {Terashima}}, \
  and\ \bibinfo {author} {\bibfnamefont {N.}~\bibnamefont {Kukubo}},\
  }\href@noop {} {\bibfield  {journal} {\bibinfo  {journal} {Phys. Rev. B}\
  }\textbf {\bibinfo {volume} {58}},\ \bibinfo {pages} {2816} (\bibinfo {year}
  {1998})}\BibitemShut {NoStop}%
\bibitem [{\citenamefont {Parendo}\ \emph {et~al.}(2004)\citenamefont
  {Parendo}, \citenamefont {Hernandez}, \citenamefont {Bhattacharya},\ and\
  \citenamefont {Goldman}}]{Parendo}%
  \BibitemOpen
  \bibfield  {author} {\bibinfo {author} {\bibfnamefont {K.~A.}\ \bibnamefont
  {Parendo}}, \bibinfo {author} {\bibfnamefont {L.~M.}\ \bibnamefont
  {Hernandez}}, \bibinfo {author} {\bibfnamefont {A.}~\bibnamefont
  {Bhattacharya}}, \ and\ \bibinfo {author} {\bibfnamefont {A.~M.}\
  \bibnamefont {Goldman}},\ }\href@noop {} {\bibfield  {journal} {\bibinfo
  {journal} {Phys. Rev. B}\ }\textbf {\bibinfo {volume} {70}},\ \bibinfo
  {pages} {212510} (\bibinfo {year} {2004})}\BibitemShut {NoStop}%
\bibitem [{\citenamefont {{Stewart Jr.}}\ \emph {et~al.}(2007)\citenamefont
  {{Stewart Jr.}}, \citenamefont {Yin}, \citenamefont {Xu},\ and\ \citenamefont
  {{Valles Jr.}}}]{Stewart}%
  \BibitemOpen
  \bibfield  {author} {\bibinfo {author} {\bibfnamefont {M.~D.}\ \bibnamefont
  {{Stewart Jr.}}}, \bibinfo {author} {\bibfnamefont {A.}~\bibnamefont {Yin}},
  \bibinfo {author} {\bibfnamefont {J.~M.}\ \bibnamefont {Xu}}, \ and\ \bibinfo
  {author} {\bibfnamefont {J.~M.}\ \bibnamefont {{Valles Jr.}}},\ }\href@noop
  {} {\bibfield  {journal} {\bibinfo  {journal} {Science}\ }\textbf {\bibinfo
  {volume} {318}},\ \bibinfo {pages} {1273} (\bibinfo {year}
  {2007})}\BibitemShut {NoStop}%
\bibitem [{\citenamefont {Sac\'{e}p\'{e}}\ \emph {et~al.}(2011)\citenamefont
  {Sac\'{e}p\'{e}} \emph {et~al.}}]{sSacepe2}%
  \BibitemOpen
  \bibfield  {author} {\bibinfo {author} {\bibfnamefont {B.}~\bibnamefont
  {Sac\'{e}p\'{e}}} \emph {et~al.},\ }\href@noop {} {\bibfield  {journal}
  {\bibinfo  {journal} {Nat. Phys.}\ }\textbf {\bibinfo {volume} {7}},\
  \bibinfo {pages} {239} (\bibinfo {year} {2011})}\BibitemShut {NoStop}%
\bibitem [{\citenamefont {Sac\'{e}p\'{e}}\ \emph {et~al.}(2010)\citenamefont
  {Sac\'{e}p\'{e}} \emph {et~al.}}]{sSacepe1}%
  \BibitemOpen
  \bibfield  {author} {\bibinfo {author} {\bibfnamefont {B.}~\bibnamefont
  {Sac\'{e}p\'{e}}} \emph {et~al.},\ }\href@noop {} {\bibfield  {journal}
  {\bibinfo  {journal} {Nat. Commun.}\ }\textbf {\bibinfo {volume} {1}},\
  \bibinfo {pages} {140} (\bibinfo {year} {2010})}\BibitemShut {NoStop}%
\bibitem [{\citenamefont {Chand}\ \emph {et~al.}(2012)\citenamefont {Chand}
  \emph {et~al.}}]{sChand}%
  \BibitemOpen
  \bibfield  {author} {\bibinfo {author} {\bibfnamefont {M.}~\bibnamefont
  {Chand}} \emph {et~al.},\ }\href@noop {} {\bibfield  {journal} {\bibinfo
  {journal} {Phys. Rev. B}\ }\textbf {\bibinfo {volume} {85}},\ \bibinfo
  {pages} {014508} (\bibinfo {year} {2012})}\BibitemShut {NoStop}%
\bibitem [{\citenamefont {Ephron}\ \emph {et~al.}(1996)\citenamefont {Ephron},
  \citenamefont {Yazdani}, \citenamefont {Kapitulnik},\ and\ \citenamefont
  {Beasley}}]{EYKB}%
  \BibitemOpen
  \bibfield  {author} {\bibinfo {author} {\bibfnamefont {D.}~\bibnamefont
  {Ephron}}, \bibinfo {author} {\bibfnamefont {A.}~\bibnamefont {Yazdani}},
  \bibinfo {author} {\bibfnamefont {A.}~\bibnamefont {Kapitulnik}}, \ and\
  \bibinfo {author} {\bibfnamefont {M.~R.}\ \bibnamefont {Beasley}},\
  }\href@noop {} {\bibfield  {journal} {\bibinfo  {journal} {Phys. Rev. Lett.}\
  }\textbf {\bibinfo {volume} {76}},\ \bibinfo {pages} {1529} (\bibinfo {year}
  {1996})}\BibitemShut {NoStop}%
\bibitem [{\citenamefont {Mason}\ and\ \citenamefont
  {Kapitulnik}(1999)}]{Mason}%
  \BibitemOpen
  \bibfield  {author} {\bibinfo {author} {\bibfnamefont {N.}~\bibnamefont
  {Mason}}\ and\ \bibinfo {author} {\bibfnamefont {A.}~\bibnamefont
  {Kapitulnik}},\ }\href@noop {} {\bibfield  {journal} {\bibinfo  {journal}
  {Phys. Rev. Lett.}\ }\textbf {\bibinfo {volume} {82}},\ \bibinfo {pages}
  {5341} (\bibinfo {year} {1999})}\BibitemShut {NoStop}%
\bibitem [{\citenamefont {Qin}\ \emph {et~al.}(2006)\citenamefont {Qin},
  \citenamefont {Vicente},\ and\ \citenamefont {Yoon}}]{Yoon2}%
  \BibitemOpen
  \bibfield  {author} {\bibinfo {author} {\bibfnamefont {Y.}~\bibnamefont
  {Qin}}, \bibinfo {author} {\bibfnamefont {C.~L.}\ \bibnamefont {Vicente}}, \
  and\ \bibinfo {author} {\bibfnamefont {J.}~\bibnamefont {Yoon}},\ }\href@noop
  {} {\bibfield  {journal} {\bibinfo  {journal} {Phys. Rev. B}\ }\textbf
  {\bibinfo {volume} {73}},\ \bibinfo {pages} {100505(R)} (\bibinfo {year}
  {2006})}\BibitemShut {NoStop}%
\bibitem [{\citenamefont {Seo}\ \emph {et~al.}(2006)\citenamefont {Seo},
  \citenamefont {Qin}, \citenamefont {Vicente}, \citenamefont {Choi},\ and\
  \citenamefont {Yoon}}]{Yoon1}%
  \BibitemOpen
  \bibfield  {author} {\bibinfo {author} {\bibfnamefont {Y.}~\bibnamefont
  {Seo}}, \bibinfo {author} {\bibfnamefont {Y.}~\bibnamefont {Qin}}, \bibinfo
  {author} {\bibfnamefont {C.~L.}\ \bibnamefont {Vicente}}, \bibinfo {author}
  {\bibfnamefont {K.~S.}\ \bibnamefont {Choi}}, \ and\ \bibinfo {author}
  {\bibfnamefont {J.}~\bibnamefont {Yoon}},\ }\href@noop {} {\bibfield
  {journal} {\bibinfo  {journal} {Phys. Rev. Lett.}\ }\textbf {\bibinfo
  {volume} {97}},\ \bibinfo {pages} {057005} (\bibinfo {year}
  {2006})}\BibitemShut {NoStop}%
\bibitem [{\citenamefont {Vicente}\ \emph {et~al.}(2006)\citenamefont
  {Vicente}, \citenamefont {Qin},\ and\ \citenamefont {Yoon}}]{Yoon3}%
  \BibitemOpen
  \bibfield  {author} {\bibinfo {author} {\bibfnamefont {C.~L.}\ \bibnamefont
  {Vicente}}, \bibinfo {author} {\bibfnamefont {Y.}~\bibnamefont {Qin}}, \ and\
  \bibinfo {author} {\bibfnamefont {J.}~\bibnamefont {Yoon}},\ }\href@noop {}
  {\bibfield  {journal} {\bibinfo  {journal} {Phys. Rev. B}\ }\textbf {\bibinfo
  {volume} {74}},\ \bibinfo {pages} {100507(R)} (\bibinfo {year}
  {2006})}\BibitemShut {NoStop}%
\bibitem [{\citenamefont {Li}\ \emph {et~al.}(2010)\citenamefont {Li},
  \citenamefont {Vicente},\ and\ \citenamefont {Yoon}}]{Yize1}%
  \BibitemOpen
  \bibfield  {author} {\bibinfo {author} {\bibfnamefont {Y.}~\bibnamefont
  {Li}}, \bibinfo {author} {\bibfnamefont {C.~L.}\ \bibnamefont {Vicente}}, \
  and\ \bibinfo {author} {\bibfnamefont {J.}~\bibnamefont {Yoon}},\ }\href@noop
  {} {\bibfield  {journal} {\bibinfo  {journal} {Phys. Rev. B}\ }\textbf
  {\bibinfo {volume} {81}},\ \bibinfo {pages} {020505(R)} (\bibinfo {year}
  {2010})}\BibitemShut {NoStop}%
\bibitem [{\citenamefont {Butko}\ and\ \citenamefont {Adams}(2001)}]{Butko}%
  \BibitemOpen
  \bibfield  {author} {\bibinfo {author} {\bibfnamefont {V.~Y.}\ \bibnamefont
  {Butko}}\ and\ \bibinfo {author} {\bibfnamefont {P.~W.}\ \bibnamefont
  {Adams}},\ }\href@noop {} {\bibfield  {journal} {\bibinfo  {journal}
  {Nature}\ }\textbf {\bibinfo {volume} {409}},\ \bibinfo {pages} {161}
  (\bibinfo {year} {2001})}\BibitemShut {NoStop}%
\bibitem [{\citenamefont {Baturina}\ \emph {et~al.}(2007)\citenamefont
  {Baturina}, \citenamefont {Strunk}, \citenamefont {Baklanov},\ and\
  \citenamefont {Satta}}]{Baturina}%
  \BibitemOpen
  \bibfield  {author} {\bibinfo {author} {\bibfnamefont {T.~I.}\ \bibnamefont
  {Baturina}}, \bibinfo {author} {\bibfnamefont {C.}~\bibnamefont {Strunk}},
  \bibinfo {author} {\bibfnamefont {M.~R.}\ \bibnamefont {Baklanov}}, \ and\
  \bibinfo {author} {\bibfnamefont {A.}~\bibnamefont {Satta}},\ }\href@noop {}
  {\bibfield  {journal} {\bibinfo  {journal} {Phys. Rev. Lett.}\ }\textbf
  {\bibinfo {volume} {98}},\ \bibinfo {pages} {127003} (\bibinfo {year}
  {2007})}\BibitemShut {NoStop}%
\bibitem [{\citenamefont {Dalidovich}\ and\ \citenamefont
  {Phillips}(2002)}]{DP}%
  \BibitemOpen
  \bibfield  {author} {\bibinfo {author} {\bibfnamefont {D.}~\bibnamefont
  {Dalidovich}}\ and\ \bibinfo {author} {\bibfnamefont {P.}~\bibnamefont
  {Phillips}},\ }\href@noop {} {\bibfield  {journal} {\bibinfo  {journal}
  {Phys. Rev. Lett.}\ }\textbf {\bibinfo {volume} {89}},\ \bibinfo {pages}
  {027001} (\bibinfo {year} {2002})}\BibitemShut {NoStop}%
\bibitem [{\citenamefont {Wu}\ and\ \citenamefont {Phillips}(2006)}]{WP}%
  \BibitemOpen
  \bibfield  {author} {\bibinfo {author} {\bibfnamefont {J.}~\bibnamefont
  {Wu}}\ and\ \bibinfo {author} {\bibfnamefont {P.}~\bibnamefont {Phillips}},\
  }\href@noop {} {\bibfield  {journal} {\bibinfo  {journal} {Phys. Rev. B}\
  }\textbf {\bibinfo {volume} {73}},\ \bibinfo {pages} {214507} (\bibinfo
  {year} {2006})}\BibitemShut {NoStop}%
\bibitem [{\citenamefont {Galitski}\ \emph {et~al.}(2005)\citenamefont
  {Galitski}, \citenamefont {Refael}, \citenamefont {Fisher},\ and\
  \citenamefont {Senthil}}]{Galitski}%
  \BibitemOpen
  \bibfield  {author} {\bibinfo {author} {\bibfnamefont {V.~M.}\ \bibnamefont
  {Galitski}}, \bibinfo {author} {\bibfnamefont {G.}~\bibnamefont {Refael}},
  \bibinfo {author} {\bibfnamefont {M.~P.~A.}\ \bibnamefont {Fisher}}, \ and\
  \bibinfo {author} {\bibfnamefont {T.}~\bibnamefont {Senthil}},\ }\href@noop
  {} {\bibfield  {journal} {\bibinfo  {journal} {Phys. Rev. Lett.}\ }\textbf
  {\bibinfo {volume} {95}},\ \bibinfo {pages} {077002} (\bibinfo {year}
  {2005})}\BibitemShut {NoStop}%
\bibitem [{\citenamefont {Shimshoni}\ \emph {et~al.}(1998)\citenamefont
  {Shimshoni}, \citenamefont {Auerbach},\ and\ \citenamefont
  {Kapitulnik}}]{SAK}%
  \BibitemOpen
  \bibfield  {author} {\bibinfo {author} {\bibfnamefont {E.}~\bibnamefont
  {Shimshoni}}, \bibinfo {author} {\bibfnamefont {A.}~\bibnamefont {Auerbach}},
  \ and\ \bibinfo {author} {\bibfnamefont {A.}~\bibnamefont {Kapitulnik}},\
  }\href@noop {} {\bibfield  {journal} {\bibinfo  {journal} {Phys. Rev. Lett.}\
  }\textbf {\bibinfo {volume} {80}},\ \bibinfo {pages} {3352} (\bibinfo {year}
  {1998})}\BibitemShut {NoStop}%
\bibitem [{\citenamefont {Spivak}\ \emph {et~al.}(2008)\citenamefont {Spivak},
  \citenamefont {Oreto},\ and\ \citenamefont {Kivelson}}]{SOK}%
  \BibitemOpen
  \bibfield  {author} {\bibinfo {author} {\bibfnamefont {B.}~\bibnamefont
  {Spivak}}, \bibinfo {author} {\bibfnamefont {P.}~\bibnamefont {Oreto}}, \
  and\ \bibinfo {author} {\bibfnamefont {S.~A.}\ \bibnamefont {Kivelson}},\
  }\href@noop {} {\bibfield  {journal} {\bibinfo  {journal} {Phys. Rev. B}\
  }\textbf {\bibinfo {volume} {77}},\ \bibinfo {pages} {214523} (\bibinfo
  {year} {2008})}\BibitemShut {NoStop}%
\bibitem [{\citenamefont {Vinokur}\ \emph {et~al.}(2008)\citenamefont {Vinokur}
  \emph {et~al.}}]{sVinokur}%
  \BibitemOpen
  \bibfield  {author} {\bibinfo {author} {\bibfnamefont {V.~M.}\ \bibnamefont
  {Vinokur}} \emph {et~al.},\ }\href@noop {} {\bibfield  {journal} {\bibinfo
  {journal} {Nature}\ }\textbf {\bibinfo {volume} {452}},\ \bibinfo {pages}
  {613} (\bibinfo {year} {2008})}\BibitemShut {NoStop}%
\bibitem [{\citenamefont {Sambandamurthy}\ \emph {et~al.}(2005)\citenamefont
  {Sambandamurthy} \emph {et~al.}}]{sSambandamurthy2}%
  \BibitemOpen
  \bibfield  {author} {\bibinfo {author} {\bibfnamefont {G.}~\bibnamefont
  {Sambandamurthy}} \emph {et~al.},\ }\href@noop {} {\bibfield  {journal}
  {\bibinfo  {journal} {Phys. Rev. Lett.}\ }\textbf {\bibinfo {volume} {94}},\
  \bibinfo {pages} {017003} (\bibinfo {year} {2005})}\BibitemShut {NoStop}%
\bibitem [{\citenamefont {Yazdani}\ and\ \citenamefont
  {Kapitulnik}(1995)}]{Yazdani}%
  \BibitemOpen
  \bibfield  {author} {\bibinfo {author} {\bibfnamefont {A.}~\bibnamefont
  {Yazdani}}\ and\ \bibinfo {author} {\bibfnamefont {A.}~\bibnamefont
  {Kapitulnik}},\ }\href@noop {} {\bibfield  {journal} {\bibinfo  {journal}
  {Phys. Rev. Lett.}\ }\textbf {\bibinfo {volume} {74}},\ \bibinfo {pages}
  {3037} (\bibinfo {year} {1995})}\BibitemShut {NoStop}%
\bibitem [{\citenamefont {Ovadia}\ \emph {et~al.}(2009)\citenamefont {Ovadia},
  \citenamefont {Sac\'{e}p\'{e}},\ and\ \citenamefont {Shahar}}]{Ovadia}%
  \BibitemOpen
  \bibfield  {author} {\bibinfo {author} {\bibfnamefont {M.}~\bibnamefont
  {Ovadia}}, \bibinfo {author} {\bibfnamefont {B.}~\bibnamefont
  {Sac\'{e}p\'{e}}}, \ and\ \bibinfo {author} {\bibfnamefont {D.}~\bibnamefont
  {Shahar}},\ }\href@noop {} {\bibfield  {journal} {\bibinfo  {journal} {Phys.
  Rev. Lett.}\ }\textbf {\bibinfo {volume} {102}},\ \bibinfo {pages} {176802}
  (\bibinfo {year} {2009})}\BibitemShut {NoStop}%
\bibitem [{\citenamefont {Altshuler}\ \emph {et~al.}(2009)\citenamefont
  {Altshuler}, \citenamefont {Kravtsov}, \citenamefont {Lerner},\ and\
  \citenamefont {Aleiner}}]{Altshuler}%
  \BibitemOpen
  \bibfield  {author} {\bibinfo {author} {\bibfnamefont {B.~L.}\ \bibnamefont
  {Altshuler}}, \bibinfo {author} {\bibfnamefont {V.~E.}\ \bibnamefont
  {Kravtsov}}, \bibinfo {author} {\bibfnamefont {I.~V.}\ \bibnamefont
  {Lerner}}, \ and\ \bibinfo {author} {\bibfnamefont {I.~L.}\ \bibnamefont
  {Aleiner}},\ }\href@noop {} {\bibfield  {journal} {\bibinfo  {journal} {Phys.
  Rev. Lett.}\ }\textbf {\bibinfo {volume} {102}},\ \bibinfo {pages} {176803}
  (\bibinfo {year} {2009})}\BibitemShut {NoStop}%
\bibitem [{\citenamefont {Matveev}\ \emph {et~al.}(1995)\citenamefont {Matveev}
  \emph {et~al.}}]{sMatveev}%
  \BibitemOpen
  \bibfield  {author} {\bibinfo {author} {\bibfnamefont {K.~A.}\ \bibnamefont
  {Matveev}} \emph {et~al.},\ }\href@noop {} {\bibfield  {journal} {\bibinfo
  {journal} {Phys. Rev. B}\ }\textbf {\bibinfo {volume} {52}},\ \bibinfo
  {pages} {5289} (\bibinfo {year} {1995})}\BibitemShut {NoStop}%
\bibitem [{\citenamefont {Mertes}\ \emph {et~al.}(1999)\citenamefont {Mertes}
  \emph {et~al.}}]{sMertes}%
  \BibitemOpen
  \bibfield  {author} {\bibinfo {author} {\bibfnamefont {K.~M.}\ \bibnamefont
  {Mertes}} \emph {et~al.},\ }\href@noop {} {\bibfield  {journal} {\bibinfo
  {journal} {Phys. Rev. B}\ }\textbf {\bibinfo {volume} {60}},\ \bibinfo
  {pages} {R5093} (\bibinfo {year} {1999})}\BibitemShut {NoStop}%
\bibitem [{\citenamefont {Klich}()}]{Klich}%
  \BibitemOpen
  \bibfield  {author} {\bibinfo {author} {\bibfnamefont {I.}~\bibnamefont
  {Klich}},\ }\href@noop {} {}\bibinfo {howpublished} {private
  communication}\BibitemShut {NoStop}%
\bibitem [{\citenamefont {Dalidovich}()}]{Dalidovich}%
  \BibitemOpen
  \bibfield  {author} {\bibinfo {author} {\bibfnamefont {D.}~\bibnamefont
  {Dalidovich}},\ }\href@noop {} {}\bibinfo {howpublished} {private
  communication}\BibitemShut {NoStop}%
\end{thebibliography}%

\end{document}